\documentclass[journal]{IEEEtran}
\usepackage{graphicx}

\begin{document}
\title{DQM4HEP - A Generic Online Monitor for Particle Physics Experiments}
%
%

\author{C. Chavez-Barajas, T. Coates and F. Salvatore~\IEEEmembership{(University of Sussex - Department of Physics and Astronomy, United Kingdom)} \\
  D. Cussans~\IEEEmembership{(University of Bristol, United Kingdom)} \\
  R. \'{E}te~\IEEEmembership{(Deutsches Elektronen-Synchrotron - DESY, Germany)} \\
  A. Irles~\IEEEmembership{(Laboratoire de l'Acc\'el\'erateur Lin\'eaire, Centre Scientifique
  d'Orsay, Universit\'e de Paris-Sud XI, 
  CNRS/IN2P3, France)}\\
  L. Mirabito~\IEEEmembership{(Universit\'e Lyon et Institut de Physique Nucléaire de Lyon - IPNL, CNRS/IN2P3, France)}\\
  A. Pingault~\IEEEmembership{(Ghent Univiersity, Belgium)\\}
  M. Wing~\IEEEmembership{(University College London, United Kingdom)}
  }


\maketitle
\pagestyle{empty}
\thispagestyle{empty}

\begin{abstract}
  There is currently a lot of activity in R\&D for future collider experiments. Multiple detector prototypes are being tested, each one with slightly different requirements regarding the format of the data to be analysed. This has generated a variety of ad-hoc solutions for data acquisition and online data monitoring. We present a generic C++11 online monitoring framework called DQM4HEP, which is designed for use as a generic online monitor for particle physics experiments, ranging from small tabletop experiments to large multi-detector testbeams, such as those currently ongoing/planned at the DESY II or CERN SPS beamlines. We present results obtained using DQM4HEP at several testbeams where the CALICE AHCAL, SDHCAL and SiWECAL detector prototypes have been tested. During these testbeams, online analysis using DQM4HEP's framework has been developed and used. We also present the currently ongoing work to integrate DQM4HEP within the EUDAQ tool. EUDAQ is a tool for common and generic data acquisition within the AIDA-2020 collaboration. This will allow these two frameworks to work together as a generic and complete DAQ and monitoring system for any type of detector prototype tested on beam tests, which is one of the goals of the AIDA-2020 project.
\end{abstract}


\section{Introduction}
There is currently a lot of activity in R\&D for future collider experiments to succeed the Large Hadron Collider, such as the International Linear Collider, Compact Linear Collider, and Future Circular Collider. Each of these will require advanced, next-generation detector technologies that must be tested extensively during development to ensure that they are capable of the sensitivities necessary to meet the physics goals of these colliders. The R\&D projects for these detectors and subdetector components are well underway, and many are currently in testing phases at beamlines around the world. 

The natural tendency is for each team to set their own standards, developing software solutions custom-tailored for their detector and development needs. In the past, this has generated a variety of ad-hoc solutions for data acquisition and online data monitoring, many of which cannot be applied outside of their original intended scope. By developing tools which are designed from the beginning to be used for many different applications, the amount of effort and development time necessary to create data acquisition and monitoring setups can be reduced significantly, simplifying and speeding up planning and deployment of physics testbeams and allowing more science to be done faster.

The AIDA-2020 project is an EU-funded project for advancing research and development infrastructure and technologies for particle physics detector development and testing, comprising 24 member countries and lead by CERN. The project is split into Work Packages; Work Package 5 is "Data acquisition system for beamtests", aiming to develop hardware and software to improve the infrastructure and tools available for testing new detector components in beams, especially for testbeams involving more than onedetector component.
The difficulty of this task is compounded by the various different detector types; different event
data models, geometries, integration times, etc.make combining data from detector components
difficult. The goal of common data acquisition is to meet this challenge by making portable
software, reducing or eliminating the work of developing DAQ systems.
The Data Quality Monitoring for High-Energy Physics framework (DQM4HEP) aims to fulfil the needs of Task 5.4: development of data quality and slow control monitoring.
Within the AIDA-2020 project, the DQM4HEP framework has been developed as a generic
online monitoring and data quality monitoring tool to meet these requirements,
allowing testbeam operators and shifters to focus on the physics goals of testbeams
and less on software engineering and integration issues.

\section{The DQM4HEP framework}
DQM4HEP is an online monitoring and data quality monitoring tool developed for physics testbeams for high-energy and particle physics. It is designed to be able to fulfil the requirements of monitoring for physics testbeams in a generic way. The structure of the program allows for independent components of the framework to be used, not used, or exchanged, by isolating each function of the program into specific and independent processes. The components that are specific to particular users -- the analysis and standalone modules -- are written in standard C++ code, meaning they are capable of performing any data unpacking, processing or analysis that is necessary. The framework then handles packaging this information in a useful way and networking to transmit it to where it is needed, meaning that the user does not have to worry about the mechanics of data storage, serialisation or transmission. It also means that the framework does not need special rules for handling particular datatypes, allowing it to handle \emph{anything} that can be packed into, decoded from, and accessed by normal C++ methods. This results in a framework that is able to deal with any kind of data, including user-defined data types, making it more flexible, portable and easily reusable. 

\subsection{Prerequisites and dependencies}
DQM4HEP is written in the C++11 standard and requires ROOT5 \cite{docROOT} for handling ROOT objects such as plots, charts and histograms. The visualisation package that contains code for the graphical user interfaces requires Qt4 \cite{docQT}. DIM is used for network communication \cite{docDIM}. There is also an optional dependency for the usage of the LCIO event data model, which is defined as the standardised filetype within AIDA-2020. If LCIO files are being used, DQM4HEP requires the LCIO software (part of the ilcSoft package \cite{docILCSoft}) in order to compile the libraries that enable serialisation of LCIO data.

\subsection{Programming paradigms and structure}
DQM4HEP is designed with genericness as its core paradigm, using processes and algorithms that are independent of data type (int, float, ROOT object, etc.). The ability to run multiple instances of each process of the framework is also key to its flexibility. This allows users to, for example, separate data that has undergone event building and data from sub-detectors, operate in online or offline modes, or distribute analysis over several networked computers to reduce computational load. 

The generic nature of the framework lies in two core features:

\begin{itemize}
\item The \textbf{Event Data Model abstraction} allows the user to define the type and structure of an event and how serialisation should be handled.
\item The \textbf{plugin system} allows the inclusion of any user-defined classes via external libraries, such as to select the serialisation process, online analysis, etc.
\end{itemize}

\begin{figure}[!t]
	\centering
	\includegraphics[width=3.0in]{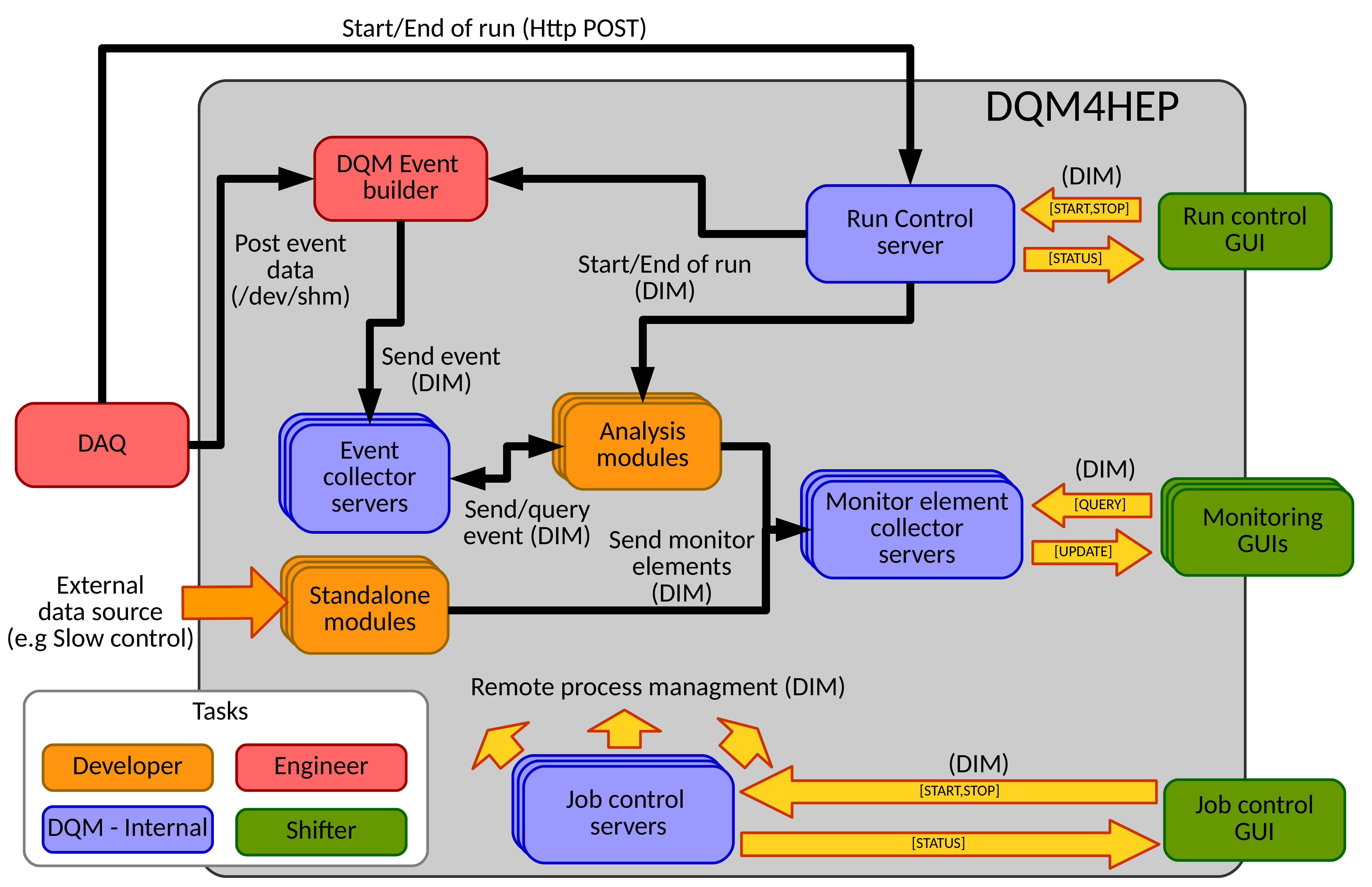}
	\caption{The global online architecture of DQM4HEP.}
	\label{fig1}
\end{figure}

The online architecture is shown in Figure \ref{fig1}. Each box represents one process of the framework and the arrows represent network communication between processes. Blue boxes are internal DQM4HEP processes that users and shifters will normally have no interaction with. Orange boxes are the "DQM modules" -- processes which must be created by the user, specific to their hardware and setup. Green boxes are interfaces used for controlling the framework or viewing the monitored quantities and properties. The colour key on the figure refers to the different roles within a team, and who is responsible for each aspect of the framework. 

The event collectors are the entrance to the framework. They collect information coming from the data acquisition system and are used to pass them to other elements in the framework. Multiple event collector processes can be run in parallel if desired. The monitor element collector receives monitor elements from the DQM modules and makes them available to the monitoring interface, and are the last exit point of data before the GUI. Again, multiple monitor element collectors can be run if desired.

There are two varieties of DQM modules:

\begin{itemize}
	\item \textbf{Analysis modules} receive events from the DAQ system via the event collector and process the event data into a form useful for monitoring, then encapsulate these as monitor elements to be sent to the monitor element collector. The analysis modules must be written for the specific usage being implemented but are produced from templates in which ordinary C++ code is used.
	\item \textbf{Standalone modules} are almost identical to analysis modules, except that they receive data from somewhere that is not the DAQ system. These are mainly used for monitoring environmental conditions.
\end{itemize}

\section{GUI and visualisation}
DQM4HEP's user interface is based on the Qt4 framework and is divided into three separate windows, which can be seen as the green boxes in Figure \ref{fig1}. These are the run control GUI, the monitoring GUI, and the job control GUI.

The run control GUI is used for starting and stopping data processing and starting new runs. It can also be used to pass parameters to analysis modules at the begin of a run. The run control GUI is optional, acting as an interface for the run control server. Some users may choose to use DQM4HEP's run control as a global run control, using it to also control their data acquisition system, while others may have a separate run control and require that DQM4HEP's run control be ``slaved'' to the central run control.

The monitoring GUI accesses the monitor elements that are collected by the monitor element collector servers. This interface is highly flexible and customisable, featuring multiple canvases. The individual monitor elements are arranged in a tree- or folder-like structure, allowing them to be organised in logical structures, such as by layer, channel, type, etc. The elements are also customisable and editable from within the UI, allowing manipulation such as zooming, scaling and fitting. Specific combinations of displayed monitor elements can be pre-set using an XML steering file, allowing complicated setups or large number of plots to be shown immediately upon startup. The update cycles of the monitor elements can also be set to either automatic or manual.

The job control GUI controls the starting and stopping of all the DQM4HEP processes, and is used to initialise and set up the framework. The job control GUI also allows operators and shifters to monitor and control the state of running processes, view or change parameters used in analysis, manually control the system while running in offline mode, and open logging files.

\section{Data quality testing}
An additional level of online monitoring is data \textit{quality} monitoring, which assesses data being received in real-time, allowing testbeam operators and shifters without detailed knowledge of the hardware to determine whether the device under test is operating as expected and fulfilling the requirements for resolution, timing, etc. Data quality monitoring (DQM) uses a variety of methods to assess the quality or ``goodness'' of data received, including ordinary statistical measures such as the mean and standard deviation, as well as slightly more advanced techniques such as the Kolmogorov-Smirnov test.

Quality tests are currently being implemented in the framework but when complete will be able to be applied to any monitor element(s) from within analysis modules. It will also be possible to run quality tests in both online and offline modes.

\begin{figure}[!t]
  \centering
  \begin{tabular}{l}
    \includegraphics[width=2.8in]{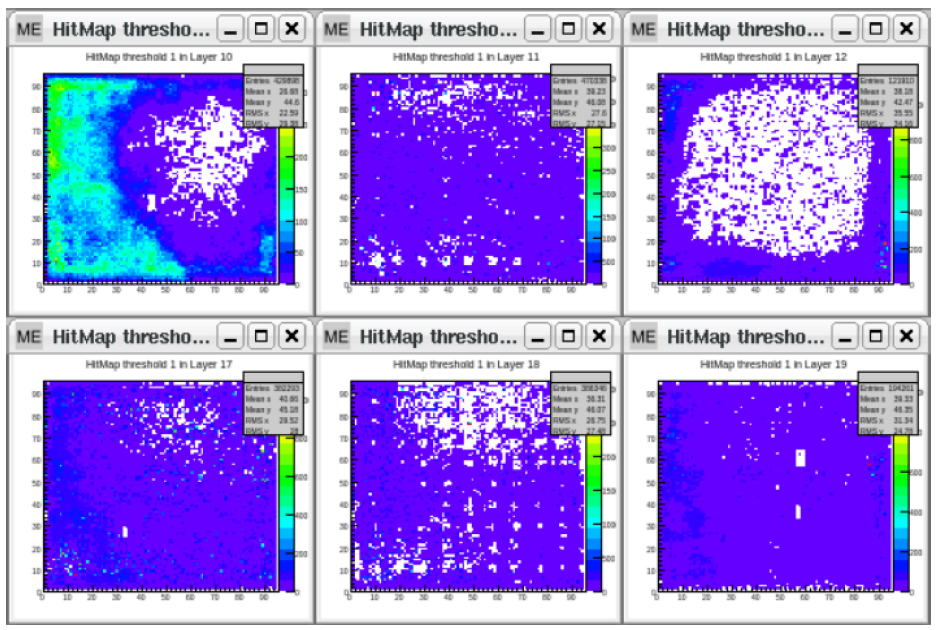}\\
    \includegraphics[width=2.8in]{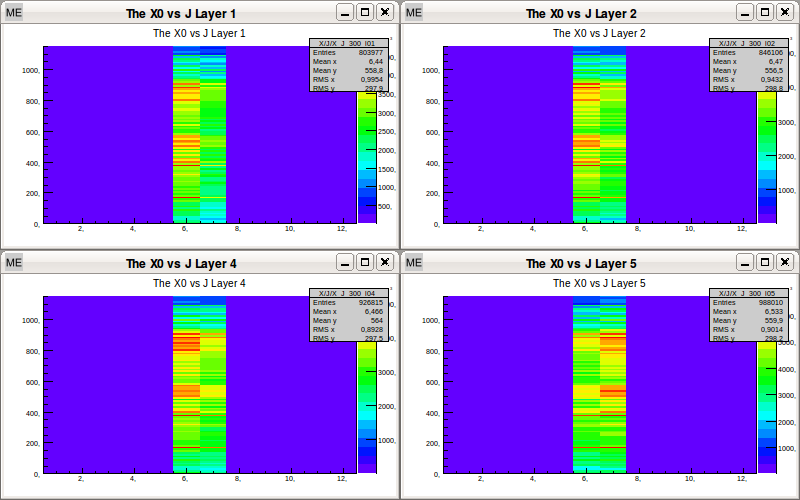}
  \end{tabular}
  \caption{Two plots produced by DQM4HEP during SDHCAL and AHCAL testbeams. Top: several hitmaps of the SDHCAL in use. Bottom: AHCAL+beam telescope correlation plots.}
  \label{fig2}
\end{figure}

\section{Implementation examples}
DQM4HEP has already been used in a number of testbeams for multiple detector prototypes, including combined testbeams. So far these have been with the AHCAL+beam telescope, and SDHCAL+SiWECAL. More information on these detectors and testbeams can be found in the references \cite{reportSDHCAL}\cite{reportAHCAL} and some examples of the framework in use can be seen in Figure \ref{fig2}.

In the two examples, interfacing with the data acquisition system was done differently. For SDHCAL testbeams, DQM4HEP was interfaced with the DAQ using shared memory (shm), allowing it to access information online. For AHCAL testbeams, the framework was used in ``nearly-online'' mode -- completed runs in LCIO format were accessed by the LCIO file service over network-attached storage as soon as an individual run file was finished. The file service can be run on files as they are being written but as events are loaded into memory, only events that were present in the file at that time were available for monitoring. Work to improve this is on-going.

During AHCAL testbeams, DQM4HEP has been and remains an incredibly useful as a tool to identify issues with the detectors during testing. For example, when new scintillator tile layers were being tested for the first time, the hitmaps allowed quick and simple visual identification of any tiles whose electronics were noisy or dead. These were plainly visible as erroneous hits for noisy channels, or gaps in the hitmap for dead channels. In SDHCAL testbeams, hitmaps were used to identify that some of the detector elements showed a lack of hits in their centres due to an overflow of incoming gas. By using hitmaps made in DQM4HEP, the problem could be identified and corrected, restarting a new run with a more stable detector.

\section{Interface with generic data acquisition software frameworks}

One of the AIDA-2020 Work Package 5 main tasks (Task 5.3) is the "Development of central DAQ software and run control". The efforts in this Task 5.3 have been focused in the developping of the EUDAQ\cite{docEUDAQ} framework.
EUDAQ was  originally designed as data acquisition software for EUDET-type beam telescopes, EUDAQ has grown to become a generic DAQ framework for other detector types.
EUDAQ is designed so that the core is flexible and portable, and all
hardware-specific components are separate and can be created, used or
ignored at the user's discretion.

The distributed process structure of EUDAQ allows individual elements to be swapped out,
saving effort and development time, compared with custom-writing an ad hoc solution that has
limited flexibility and portability. While EUDAQ has an online monitoring component,
it is not being discussed, and may be removed from future versions in favour of DQM4HEP.

EUDAQ can be used as a generic DAQ, while DQM4HEP can be used as a generic data quality
monitoring tool. Both are hardware-independent and when used in concert may form a fullyfeatured,
generic and portable DAQ/DQM system, replacing most software used during beam tests.

Development of an online linkage is underway,
which will allow EUDAQ to stream events to
DQM4HEP processes online.
Once this is completed, the combined
EUDAQ/DQM4HEP system will allow a fully-generic
DAQ and monitoring system. The only detector specific
components will be:

\begin{itemize}
\item EUDAQ Producer and DataConverterPlugin
\item Event type and serialisation method
\item Online anaylsis tasks and modules
\end{itemize}

\section{Conclusion}
With its generic and flexible programming, DQM4HEP forms a powerful and portable framework for online monitoring and data quality monitoring for physics testbeams that will allow physicists to focus on the physics goals instead of the engineering and software issues of building and deploying their own monitoring systems. Planned future work to allow DQM4HEP to interface directly with EUDAQ, a generic data acquisition system within AIDA-2020, will allow the two software frameworks to work together to form a generic data acquisition, monitoring and quality monitoring system that is capable of being used for nearly any type of detector model.

The DQM4HEP framework can be found on Github at \texttt{https://github.com/DQM4HEP}. User and technical documentation is currently in progress but any enquiries about the framework can be directed to \texttt{dqm4hep@gmail.com}.


\appendices

\section*{Acknowledgment}
This project has received funding from the European Union's Horizon 2020 Research and Innovation programme under Grant Agreement no. 654168.
The research leading to these results has received funding from the People Progr
amme (MarieCurie Actions) of the European Union’s Seventh Framework Programme (FP7/2007-2013)
under REA grant agreement n. PCOFUND-GA-2013-609102, through the PRESTIGE
programme coordinated by Campus France.



\begin{thebibliography}{1}
  	\bibitem{docROOT} R. Brun and F. Rademakers, \emph{ROOT -- An Object Oriented Data Analysis Framework}, Nucl. Inst. \& Meth. in \emph{Phys. Res. A 389}, 1997, pp. 81-86.  Available at: http://root.cern.ch
	\bibitem{docQT} Qt Company, http://www.qt.io, v4.7, 2016.
		\bibitem{docDIM} C. Gaspar et al., \emph{DIM a Portable Lightweight Package for Information Publishing Data Transfer and Inter-process Communication} presented at the Int. Conf. Computing in High Energy and Nuclear Physics, Padova, Italy, 2000)
	\bibitem{docILCSoft} F. Gaede and J. Engels, 2007. \emph{Marlin et al-A Software Framework for ILC detector R\&D}. EUDET-Report-2007-11. Available at: http://ilcsoft.desy.de/portal
	\bibitem{reportSDHCAL} CALICE collaboration, 2016. \emph{First results of the CALICE SDHCAL technological prototype}. \emph{Journal of Instrumentation}, 11(04), p.P04001. [{\tt hep-ex/160202276}]
	\bibitem{reportAHCAL} C. Adloff, Y. Karyotakis, J. Repond, A. Brandt, H. Brown, K. De, C. Medina, J. Smith, J. Li, M. Sosebee and A. White, 2010. \emph{Construction and commissioning of the CALICE analog hadron calorimeter prototype}. \emph{Journal of Instrumentation}, 5(05), p.P05004. [{\tt hep-ex/10032662}]
	\bibitem{docEUDAQ} Y. Liu. \emph{EUDAQ2 User Manual}. Retrieved October 13, 2017, from https://github.com/eudaq/eudaq 
\end{thebibliography}
%

\end{document}